\documentclass[a4paper,fleqn,usenatbib]{mnras}

\usepackage{newtxtext,newtxmath}

\usepackage[T1]{fontenc}
\usepackage{ae,aecompl}

\usepackage{graphicx}	
\usepackage{amsmath}	
\usepackage{amssymb}	
\usepackage{color}

\newcommand{{\vw}}{VW~Hyi}
\newcommand{{\suz}}{\it Suzaku}
\newcommand{{\xmm}}{\it XMM-Newton}
\newcommand{{\asc}}{\it ASCA}
\newcommand{{\gin}}{\it Ginga}
\newcommand{{\ro}}{\it ROSAT}

\title[Variation of mass accretion rate of VW Hyi]{Variation of mass accretion rate onto the white
  dwarf in the dwarf nova VW Hyi in quiescence}

\author[N. Nakaniwa et al.]{
Nozomi Nakaniwa,$^{1,2}$\thanks{E-mail: nakaniwa-nozomi@ed.tmu.ac.jp}
Takayuki Hayashi,$^{3}$
Mai Takeo,$^{1,2}$
and Manabu Ishida$^{1,2}$
\\
$^{1}$Department of Physics, Tokyo Metropolitan University, 1-1 Minami-Osawa, Hachioji, Tokyo
192-0397, Japan\\
$^{2}$The Institute of Space and Astronautical Science/JAXA, 3-1-1 Yoshinodai, Chuo-ward,
Sagamihara, Kanagawa 252-5210, Japan\\
$^{3}$NASA's Goddard Space Flight Center, 8800 Greenbelt Road, Greenbelt, MD 20771, USA
}

\date{Accepted 2019 June 29. Received 2019 June 28; in original form 2019 May 17}

\pubyear{2015}

\begin{document}
\label{firstpage}
\pagerange{\pageref{firstpage}--\pageref{lastpage}}
\maketitle

\begin{abstract}
We have analysed a series of {\suz} data and one dataset of {\xmm} of the SU~UMa type dwarf nova
{\vw} in optical quiescence. The observed spectra in the 0.2-10~keV band are moderately well
represented by multi-temperature thermal plasma emission models with a maximum temperature of
5-9~keV and bolometric luminosity of $(2.4-5.2)\times 10^{30}$ erg~s$^{-1}$. The mass accretion
rate derived from the hard X-ray spectra does not show any clear trend as a function of time since
the last superoutburst, in contradiction to theoretical predictions of the disc behaviour of a
SU~UMa type dwarf nova. The mass accretion rate, on the other hand, shows a clear declining trend with time since the last outburst (including the superoutburst). The rate of decline is of the
same order as that evaluated from the hard X-ray light curves of the other two dwarf novae SS~Cyg and
SU~UMa. The standard disc instability model, on the other hand, predicts that the mass accretion
rate should increase throughout the optically-quiescent phase. We need further observation and
theoretical consideration to resolve this discrepancy.
\end{abstract}

\begin{keywords}
stars: dwarf novae -- novae, cataclysmic variables -- X-rays: binaries -- X-rays: stars.
\end{keywords}



\section{Introduction}
\label{section:Intro}

A cataclysmic variable (CV) is a binary system composed of a Roche-lobe-filling
secondary star and a mass-accreting white dwarf \citep{2003cvs..book.....W}. The matter spilt over
the Roche Lobe forms an accretion disc around the white dwarf. The disc is basically optically thick
with an innermost temperature of order 10$^5$~K. Hence its radiation appears in the band from
infrared to ultraviolet. 
A dwarf nova (DN) is a subclass of CV in which an optical outburst occurs with a recurrence time of months to decades. 
During the outburst, the DN becomes brighter than in quiescence by a few to several magnitudes in the V-band. 
The accretion disc of the DN is bi-stable. One stable state corresponds to a low accretion-rate state in which the temperature of the outer part of the disc is low (of order 10$^3$~K) and hydrogen there is neutral. The other is a high accretion-rate state in which the outer part temperature exceeds $10^4$~K and the hydrogen is ionized. The former and the latter correspond to an optically-quiescent state and an outburst state, respectively. Since the mass accretion rate from the secondary star is intermediate, the DN shows switching behaviour between the two states due to thermal instability associated with hydrogen ionization/recombination \citep{1974PASJ...26..429O,1979PThPh..61.1307H,1981A&A...104L..10M}.

Due to lack of any coherent periodic signal, unlike in magnetic cataclysmic variables, it is believed
that the magnetic field of the white dwarf is weak, say $B \la 10^6$~G \citep{1996A&A...315..467V},
and the accretion disc reaches the white dwarf surface. If the white dwarf rotates much more slowly
than the Keplerian velocity at its surface, the accreting matter experiences high friction between
the inner accretion disc and the white dwarf surface, and as a result, is abruptly heated. This
region is referred to as the boundary layer.

Since the density of the disc is relatively low in the quiescent state, the heating efficiency
surpasses the cooling rate in the boundary layer. Consequently, the matter in the boundary layer
swells, leaves the orbital plane, and forms an optically thin thermal plasma with a temperature of
order 10$^8$~K, from which hard X-ray emission emanates. In the outburst state, on the other hand,
the cooling efficiency is in general high enough to keep the disc optically thick, and the emission
from the disc extends to EUVE region \citep{1979MNRAS.187..777P,1985ApJ...292..535P}. The transition
between these two states was clearly detected with the multi-waveband observation of SS~Cygni by
\citet{2003MNRAS.345...49W}, in which an outburst in the EUVE band was delayed from that in optical
by $\sim$1~d, and its onset coincided with an abrupt decline of the hard X-ray emission in the
2-15~keV band.
\begin{center}
\begin{table*}
\begin{minipage}{0.8\textwidth}
\caption{{\suz} observation log of {\vw} with {\suz} and {\xmm}. \label{tab:obslog}}
\begin{tabular}{ccccccc} \hline
Sequence{\rm \#} & Observation date                & Observatory & Detector     & Mode   & Exposure$^\dagger$ 
  & Intensity$^\ddagger$\\
                 & (UT)                            &             &              &        & (ks) 
  & (count s$^{-1}$)$^\S$ \\ \hline
4006009020       & 2011 Dec 29 15:19:16 - 04:20:11 & {\suz}      & XIS FI       & Normal & 18.35
  & 1.082 $\pm$ 0.008\\
                 &                                 &             & XIS BI       & Normal & 18.35
  & 0.794 $\pm$ 0.007\\
4006009030       & 2012 Feb 29 13:39:04 - 06:00:11 & {\suz}      & XIS FI       & Normal & 22.83
  & 0.928 $\pm$ 0.007\\
                 &                                 &             & XIS BI       & Normal & 22.83
  & 0.633 $\pm$ 0.006\\
4006009040       & 2012 May 02 19:29:54 - 04:00:13 & {\suz}      & XIS FI       & Normal & 18.08
  & 0.994 $\pm$ 0.008\\
                 &                                 &             & XIS BI       & Normal & 18.08
  & 0.654 $\pm$ 0.006\\ \hline
0111970301       & 2001 Oct 19 05:31:04 - 10:42:58 & {\xmm}      & MOS 1        & Normal & 18.17
  & 0.656 $\pm$ 0.006 \\
	         &                                 &             & MOS 2        & Normal & 18.17
  & 0.688 $\pm$ 0.007 \\
                 &                                 &             & PIN          & Normal & 14.45
  & 2.354 $\pm$ 0.013\\ \hline
\end{tabular}
\end{minipage}
\begin{minipage}{0.8\textwidth}
\footnotesize 
$^\dagger$ After data screening which is described in \S\ref{section:DSC}.\\ 
$^\ddagger$ In the band 0.5-10.0 keV for the XIS FI (XIS0 + XIS3) and 0.3-10.0 keV for the
XIS BI.\\
$^\S$ After background subtraction. The source and background regions are explained in
\S\ref{subsec:EoS}.
\end{minipage}
\end{table*}
\end{center}

The hot plasma in the boundary layer of the quiescent state is formed by the frictional heating of the
disc matter, is cooled via optically thin thermal plasma emission, and is finally settled onto the
white dwarf surface. Hence it is multi-temperature. 
As a matter of fact, a cooling flow model that was originally developed for clusters of
galaxies \citep{1988ASIC..229...53M,1994ARA&A..32..277F} and its slightly modified version has been successful
in fitting DN hard X-ray spectra in general
\citep{2005MNRAS.357..626B,2005ApJ...626..396P,2017PASJ...69...10W}.

VW~Hydri is a DN whose primary and secondary masses are $0.63\pm 0.15M_\odot$ and $0.11\pm
0.02M_\odot$, respectively, and the orbital period is 107.0~min
\citep{1981A&A....97..185S,2003A&A...404..301R}. There are a couple of estimations of the distance
to {\vw}: 65~pc \citep{1987MNRAS.227...23W} and $82\pm 5$~pc \citep{1996PASP..108..412B}. Recently,
however,
the {\it Gaia} collaboration \citep{2016A&A...595A...1G} published results
  of their highly accurate parallax measurements. Just inverting the parallax of {\vw} from data
  release 2 \citep[DR2,][]{2018A&A...616A...1G}, we obtained the distance to {\vw} as $54\pm 0.1$~pc.

As can be recognized from the orbital period, {\vw} is an SU~UMa-type DN which shows not only the
normal outburst but also a superoutburst. The superoutburst is brighter in the
optical band than the normal outburst by $\sim$1~mag at its peak, and lasts for $\sim$7-10~d, longer
than the normal outburst ($\sim$ a few days). During the normal outburst, the disc mass continues to
decline due to accretion onto the white dwarf, but the total mass loss per outburst is smaller than
the gain of mass between the two outbursts. Hence the disc gradually gets heavier and extends
outward. 
Once the disc grows large enough, a thermal-tidal instability occurs,
  triggered by a resonance between the orbital motion and the Keplerian motion of the disc, leading
  to the superoutburst
  \citep{1988MNRAS.232...35W,1989PASJ...41.1005O,1990PASJ...42..135H,1996PASP..108...39O}. 
  In the case of {\vw}, the normal outburst and the superoutburst occur every $\sim$30~d and $\sim$180~d,
respectively \citep{2003A&A...404..301R}.

The X-ray behaviour of {\vw} has been investigated using various
observatories. \citet{1996A&A...307..137W} observed a termination phase of an
  outburst in the 2-10~keV band with {\it Ginga}, and fitted the cooling flow model
\citep{1988ASIC..229...53M} to the combined {\gin} and {\ro} spectra. This gave a maximum
temperature of the boundary layer of $11^{+3}_{-2}$~keV. The flux in the 0.04-10~keV band was
$2.08\times 10^{-11}$ erg~cm$^{-2}$~s$^{-1}$. \citet{2003MNRAS.346.1231P} observed {\vw} with {\xmm}
during the middle of a normal outburst and the following superoutburst. The cooling flow model
results in $T_{\rm max} = 7.8\pm 0.2$~keV with a bolometric flux of $7.8\times 10^{-12}$
erg~cm$^{-2}$~s$^{-1}$. \citet{2005MNRAS.357..626B} analysed a couple of {\it ASCA} datasets of
{\vw} in the quiescent state. One was taken at the middle of the end of a superoutburst and the
following normal outburst, and the other is $\sim$17~d after a normal outburst. The intensity of the
former observation is $\sim$4 times as large as the latter. The flux of the former in the band
0.8-10~keV was $4.7\times 10^{-12}$ erg~cm$^{-2}$~s$^{-1}$ with a fixed $T_{\rm max}$ of 20~keV.

In this paper, we present results from a series of {\vw} observations carried out with {\suz} in the
optically-quiescent state. There is a prediction of the time evolution of the mass transfer rate through
the outer disc within a supercycle. The mass transfer rate suddenly increases at the onset of a
normal/super outburst and continues to decline to the end of the outburst. During the interval
period of two neighbouring outbursts, the transfer rate is predicted to increase by degrees due to
accumulation of mass from the secondary \citep[][and references
  therein]{1996PASP..108...39O}. 
On the other hand, there has been no systematic study of the mass transfer rate in the
boundary layer in relation to the supercycle phase. Since the boundary layer in the quiescent
state mainly emits hard X-rays, we have decided to analyse {\suz} data systematically to elucidate
the mass accretion history in the quiescent state. This paper is organized as follows. In \S~2, we
summarize the {\suz} observations of {\vw} in the quiescent state. Observation windows and data
selection criteria are described. We also reanalysed the {\xmm} data of {\vw} in quiescence, which
is already published by \citet{2003MNRAS.346.1231P}. In \S~3, we explain our spectral
analysis in detail. \S~4 is devoted to discussion. We primarily concentrate on the mass accretion
rate history in relation to the supercycle or the outburst cycle. Finally, we summarize our
results in \S~5.
\begin{figure*}
\centerline{
  \includegraphics[angle=270,width=0.9\textwidth]{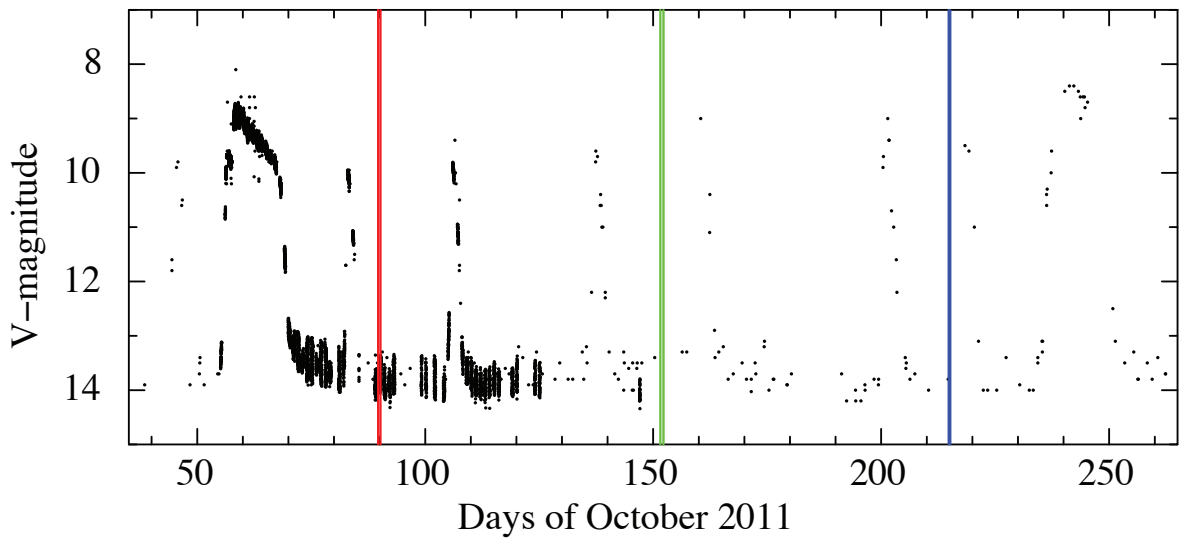}
}
  \caption{The {\suz} observation windows (the red, green and blue bands) overlaid on the optical
    light curve, which is taken from the AAVSO homepage. The abscissa shows the
      elapsed time in days since 2011 October 1st 0:00:00 (UT).}\label{fig:aavso_suzaku}
\end{figure*}

\section{Observations and data reduction}

\subsection{Suzaku Observation}

{\suz} \citep{2007PASJ...59S...1M} observed {\vw} in the quiescent state three times in
2011-2012. {\suz} is equipped with two kinds of detector systems. One is the X-ray Imaging
Spectrometer (XIS; \citealt{2007PASJ...59S..23K}), which adopts a Charge-Coupled Device (CCD). Of
the four modules XIS0 through XIS3, XIS1 is a back-illuminated (BI) CCD whereas the others are
front-illuminated (FI) CCDs. Each XIS module is mounted in the focal plane of the corresponding
X-Ray Telescope (XRT; \citealt{2007PASJ...59S...9S}) whose focal length is 4.75~m. Note, however,
that XIS2 has not been functional since 2006 November 9th, probably because of a micro-meteorite
collision. The other detector system is the Hard X-ray Detector (HXD;
\citealt{2007PASJ...59S..35T,2007PASJ...59S..53K}), which is a non-focusing hard X-ray detector
covering 10-600~keV. We do not use the HXD data, because {\vw} was too faint for spectral
analysis in the HXD band. Table~\ref{tab:obslog} summarizes the {\suz} observations of {\vw}
in the quiescent state. In Fig.~\ref{fig:aavso_suzaku} the {\suz} observation windows in the
quiescent state are shown overlaid on the optical light curve taken from the homepage of the American
Association of Variable Star Observers (AAVSO)\footnote{https://www.aavso.org}. In addition to these
quiescence data, there is one more dataset taken near the peak of the superoutburst (around day
60 in Fig.~\ref{fig:aavso_suzaku}). The observation and the analysis results are
summarized in \citet{2017PASJ...69...10W}, together with those from quiescence.

\subsection{XMM-Newton Observation}

\begin{figure*}
\centerline{
  \includegraphics[angle=270,width=0.9\textwidth]{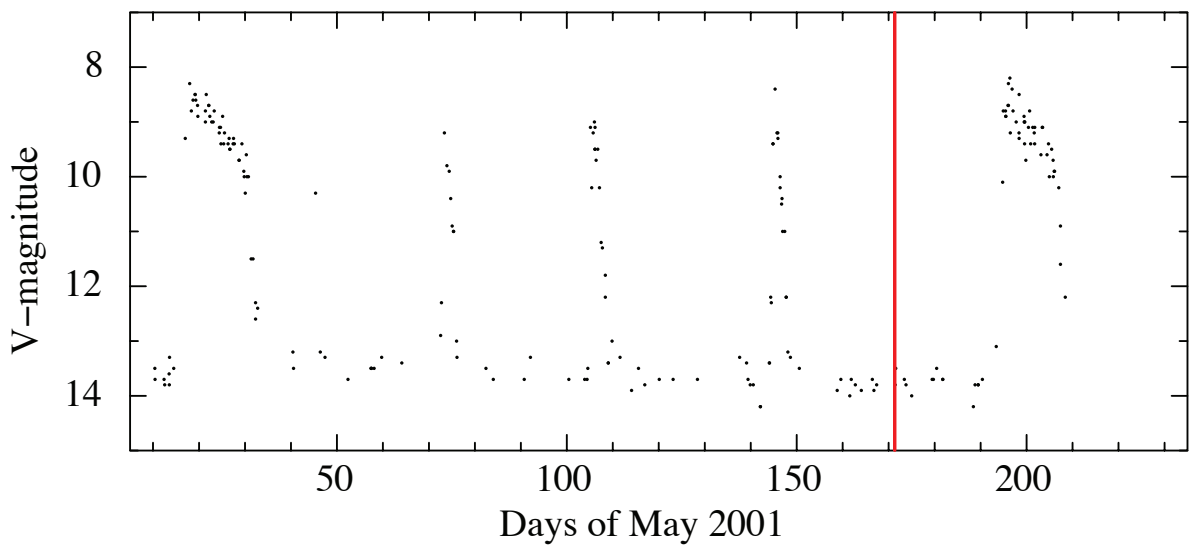}
}
  \caption{The {\xmm} observation window (the red band) overlaid on the optical light curve, which
    is taken from AAVSO homepage. The abscissa shows the elapsed time in days
      since 2001 May 1st 0:00:00 (UT).}\label{fig:aavso_xmm}
\end{figure*}

As well as {\suz}, {\xmm} \citep{2001A&A...365L...1J} has observed {\vw} in its quiescent state,
which has already been published by \citet{2003MNRAS.346.1231P}. Since the effective area of {\xmm}
is large and advantageous for spectroscopy of a faint source like {\vw}, we have decided to
incorporate these data in our analysis. The observation log of {\xmm} is also listed in
table~\ref{tab:obslog}, and the observation epoch is shown overlaid on the AAVSO light curve in
Fig.~\ref{fig:aavso_xmm}.
The observation was carried out during an epoch between a normal outburst and a following
superoutburst in 2001.

\subsection{Data selection criteria}
\label{section:DSC}

\subsubsection{\suz}

We have screened the {\suz} data with {\sc xselect} version 2.4e, which is included in the HEASOFT
package provided by NASA's Goddard Space Flight Center
(GSFC)\footnote{https://heasarc.gsfc.nasa.gov/docs/software.html}. We have mostly followed the
standard data selection criteria. We have only used data while the telemetry rate is either high or medium. We have removed bad pixels and flickering
pixels, and have selected events of grade 0, 2, 3, 4 and 6. We have discarded data taken during
spacecraft passages through the South Atlantic Anomaly and when the pointing accuracy is low. Also
discarded are data taken while the elevation of {\vw} from the night earth limb is less than
5~deg. On the other hand, the standard data screening criteria recommend not to use the data when
the target star is within 20~deg from the bright earth limb. The number, however, depends upon
attitude and orbit condition at the time of the observation. In addition to this, {\vw} lies close
to the south ecliptic pole, and hence the XIS field of view tends to be close to the earth
limb. These facts suggest that significant amounts of data would be discarded with the standard
day-earth elevation angle. Accordingly, we have drawn spectra with various trial day-earth elevation
angles and, by closely inspecting the spectra in the band 0.3-0.7~keV, where the atmospheric
Nitrogen and Oxygen lines appear, we have found that the cutoff day-earth elevation angle can be set
as small as 7~deg. The exposure times listed in table~\ref{tab:obslog} are those after applying the
new day-earth elevation threshold. As we expected, the exposure times are increased by 7-14\%.

\subsubsection{\it XMM-Newton}

To increase observation coverage over the supercycle phase, we reanalysed the {\it XMM-Newton} data of
{\vw}, which were published by \citet{2003MNRAS.346.1231P}. We have used the Science
Analysis Software (SAS) {\sc xmmsas\_20160201\_1833-15.0.0}. We have followed the standard data
selection criteria.

\section{Analysis and results}

\subsection{Extraction of spectra}
\label{subsec:EoS}

In extracting spectra of the {\suz} XIS, we have collected X-ray photons arriving in a circular
aperture with a radius of $3\farcm5$ centered on {\vw}. The background photons have been integrated
over an annular region with inner and outer radii of 4$'$ and $6\farcm5$, respectively. As for the
{\xmm} MOS and pn, the source photons have been accumulated with a circular aperture of $40"$
radius, while those of the background are from an annulus with inner and outer radii of $40''$ and
$60''$, respectively.

\begin{figure}
\centerline{
  \includegraphics[angle=0,width=0.435\textwidth]{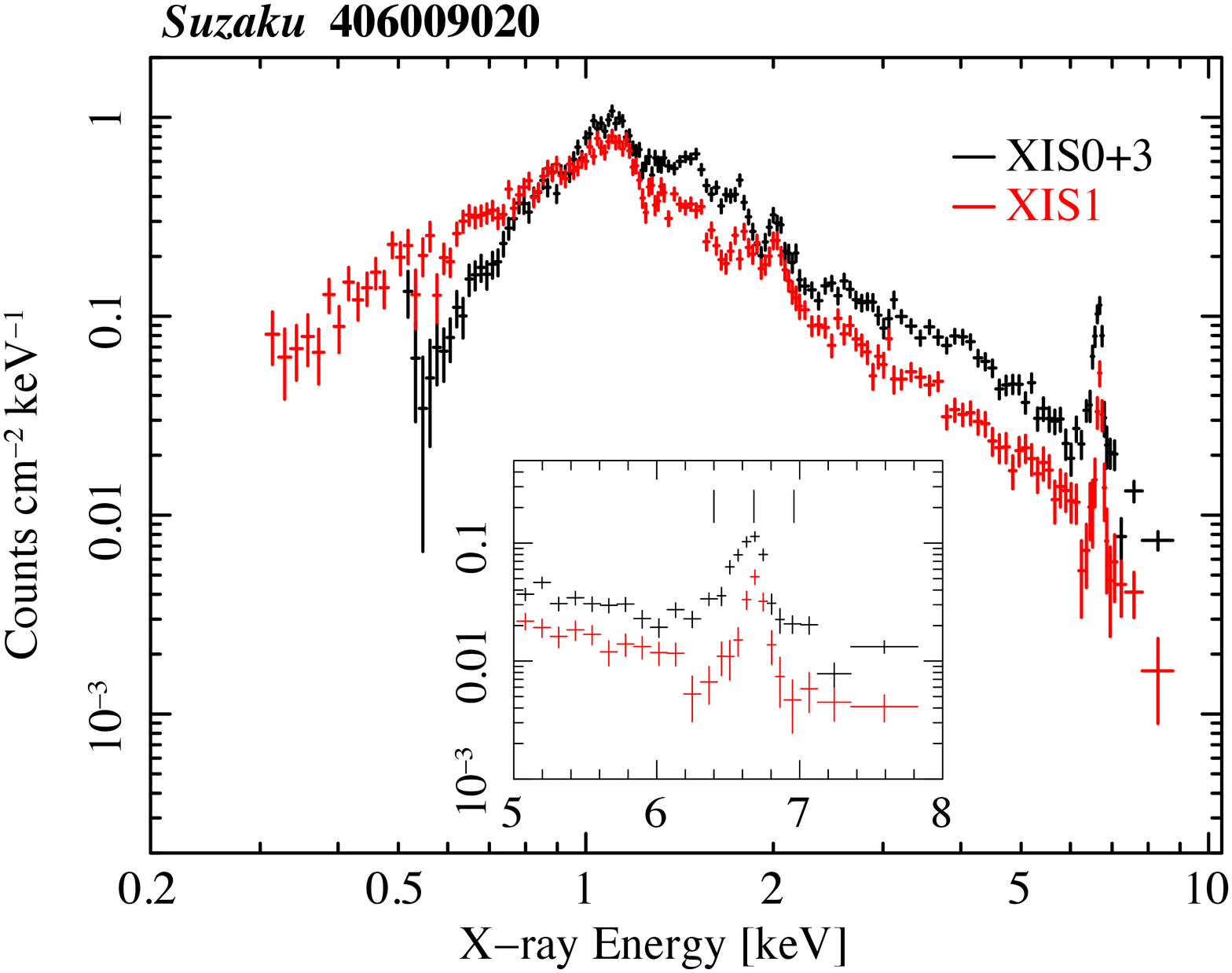}
}
\centerline{
  \includegraphics[angle=0,width=0.435\textwidth]{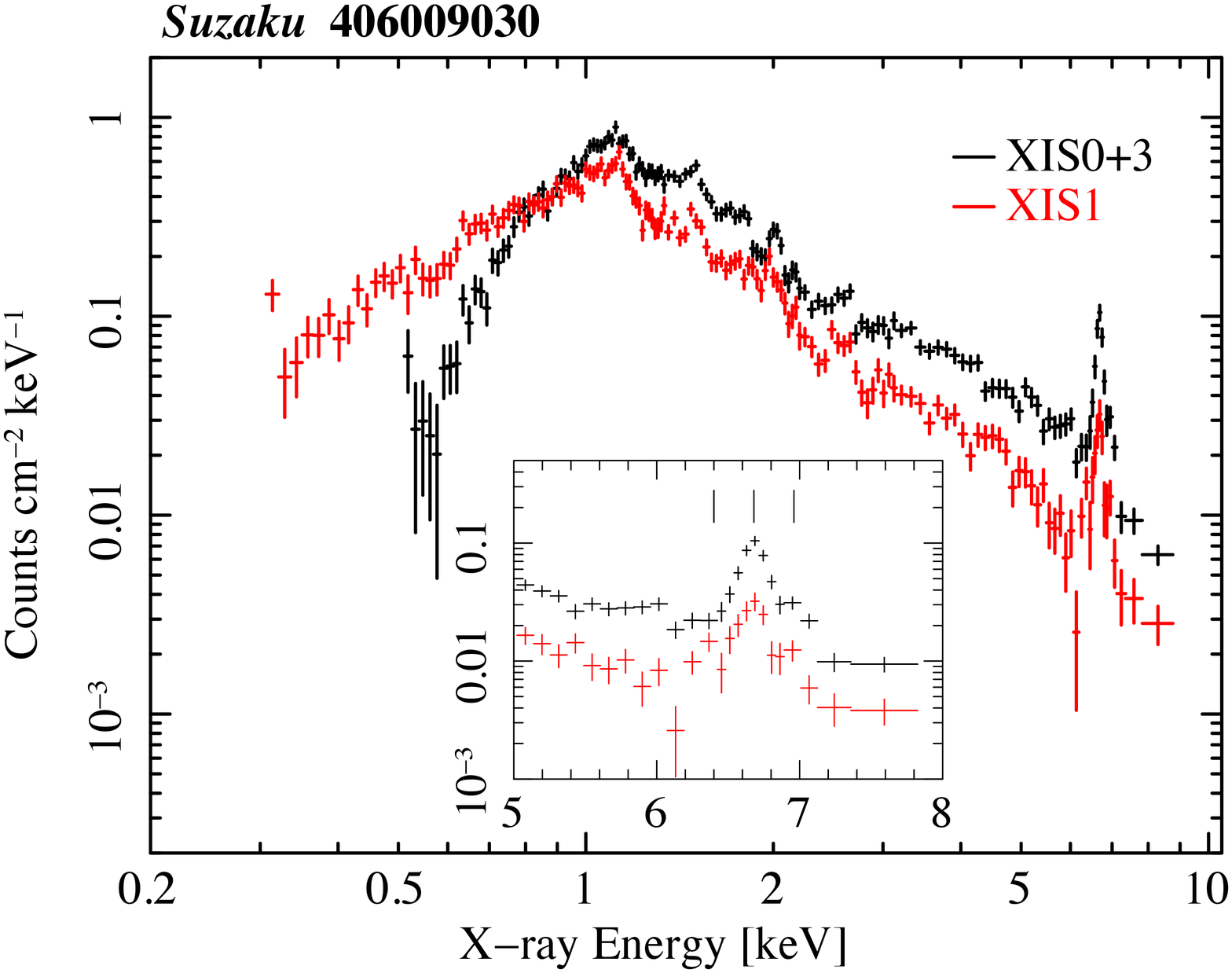}
}
\centerline{
  \includegraphics[angle=0,width=0.435\textwidth]{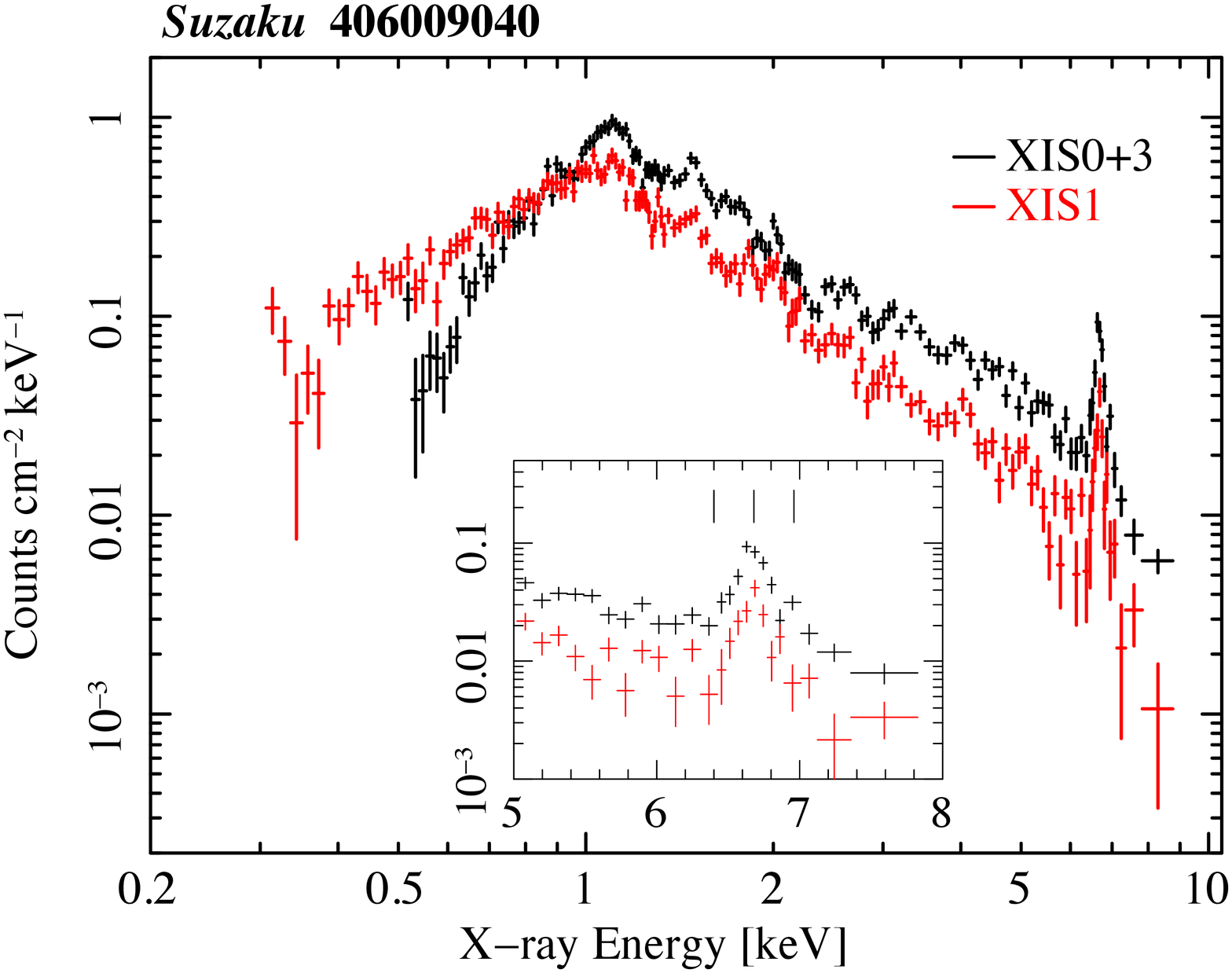}
}
\centerline{
  \includegraphics[angle=0,width=0.435\textwidth]{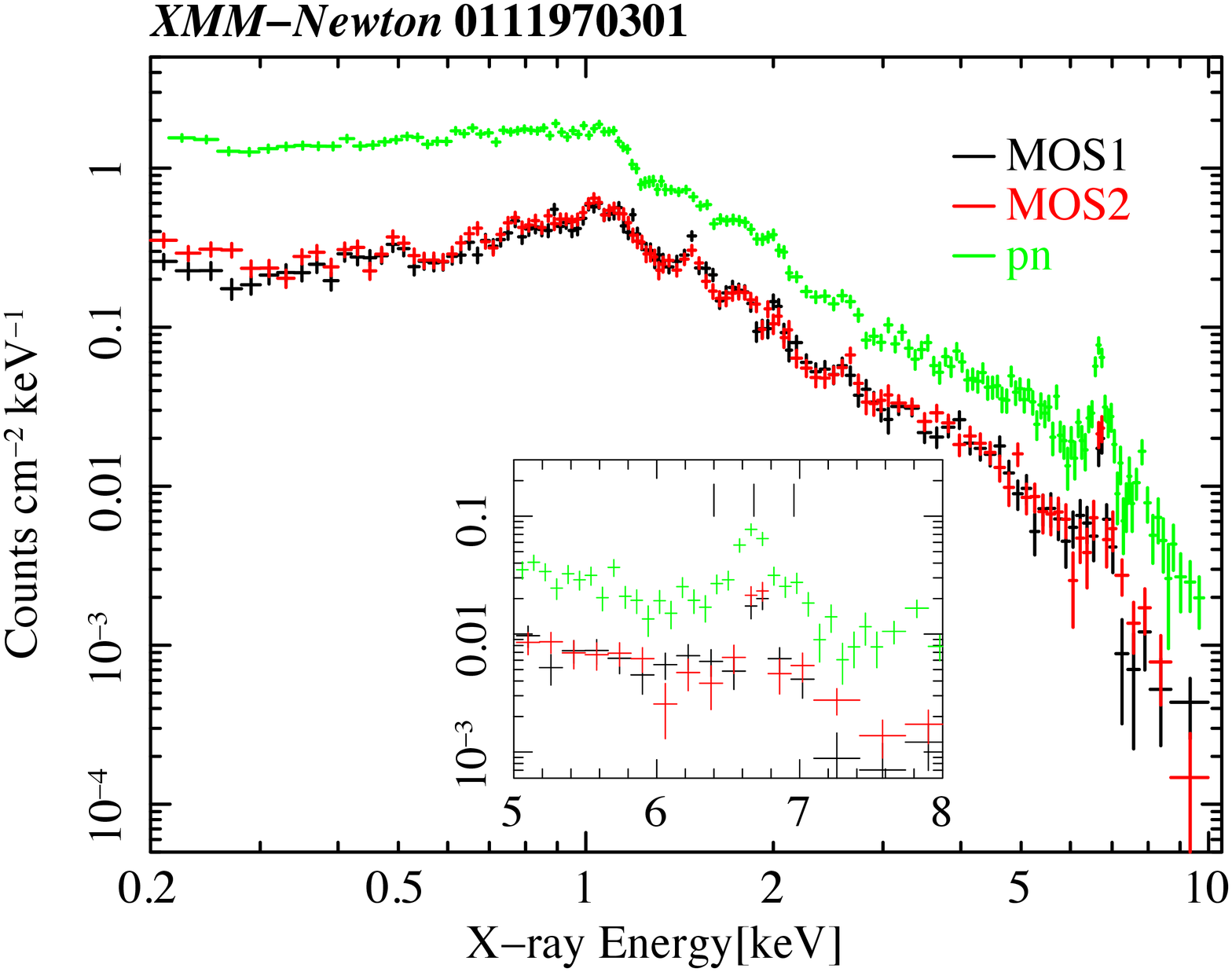}
}
\caption{{\vw} spectra of {\suz} (seq. \#406009020, 406009030, 406009040) and {\xmm}
  (seq. \#0111979301). The tick marks of the insets indicate the energies of iron K$\alpha$ emission
  lines from neutral, He-like, and hydrogenic ionization states. Note that the ordinate range of the
  {\xmm} main frame is different from that of {\suz}. \label{fig:raw_spectra}}
\end{figure}

The spectra thus obtained are shown in Fig.~\ref{fig:raw_spectra}.
The insets are blow-ups of the iron K$\alpha$ emission line energy band. Although statistics are
somewhat limited, the spectra are characterized by He-like and hydrogenic K$\alpha$ emission
lines (hereafter referred to as He$\alpha$ and Ly$\alpha$, respectively) of Mg (1.34~keV and
1.47~keV), Si (1.86~keV and 2.01~keV), S (2.45~keV and 2.62~keV) and Fe (6.68~keV and
6.96~keV). This fact strongly suggests that the X-ray emission from {\vw} originates from a thermal
plasma with multiple temperatures. As recognized from the insets, the iron K$\alpha$ emission is
dominated by the He$\alpha$ component, whereas Ly$\alpha$ is detected only marginally. This implies
that the maximum temperature of the plasma is lower than $\sim$10~keV. It is also remarkable from
the insets that the neutral iron K$\alpha$ line (6.40~keV) seems absent. This is a big contrast to
the archetypal DN SS~Cygni \citep{2009PASJ...61S..77I}.

\subsection{Response functions}

In preparation for spectral analysis, we have made redistribution matrix files
  (RMFs) of the {\suz} XIS with the software {\sc xisrmfgen} (version 2012-04-21). With the RMF thus
  made, we have then created ancillary response files (ARFs) with {\sc xissimarfgen} (version
  2010-11-05). The information of the photon-integration region is taken into account here. The RMFs
  and the ARFs thus produced are merged with {\sc marfrmf} (v3.2.6) for the three XIS modules
  separately, and the resultant response matrix files of the XIS0 and XIS3 (FI CCDs) are coadded
  with {\sc addrmf}.

We have followed a similar procedure in preparing the response matrix files for
  {\xmm} MOS and pn. We have made the RMFs of MOS1, MOS2, and pn separately with {\sc rmfgen}
  (version 2.2.1) and the ARFs with {\sc arfgen} (version 1.92), which are both included in the
  package xmmsas\_20160201\_1833-15.0.0.

\subsection{Spectral models}

In evaluating the spectra, we use the software {\sc xspec} \citep{1996ASPC..101...17A} version
12.10.0 in the HEASOFT package provided by NASA's GSFC\footnote{http://heasarc.gsfc.nasa.gov}. Based
upon the appearance of the spectra described above, as well as upon accumulated knowledge of the DNe
X-ray spectra, we here adopt the following models. Throughout this paper, we adopt $N_{\rm H} =
6\times 10^{17}$~cm$^{-2}$ \citep{1990ApJ...356..211P} as the interstellar absorption hydrogen
column density to {\vw}.

\subsubsection{Multiple temperature plasma emission models}

We have adopted the following two spectral models to represent the observed multiple temperature
plasma emission. One is the {\sc vmcflow} model. As can be deduced from its name, this model is
originally developed to model X-ray spectra of a cooling flow that was presumed to appear in galaxy
clusters \citep{1988ASIC..229...53M,1994ARA&A..32..277F}.
Probably because the heating and cooling processes that take place in the boundary layer is similar
to those in the clusters of galaxies, {\sc vmcflow} has been found to succeed in representing X-ray
spectra of DNe in general \citep{2005ApJ...626..396P,2005MNRAS.357..626B,2017PASJ...69...10W}. In
this model, the mass accretion rate $\dot{M}$ is one of the model parameters which is evaluated from
the equation
\begin{equation}
L_{\rm bol} = \frac52 \frac{\dot{M}}{\mu m_{\rm H}} k_{\rm B}T_{\rm max},
\end{equation}
where $T_{\rm max}$ is the maximum temperature of the plasma, which is again one of the model
parameters, and $L_{\rm bol}$ is the bolometric luminosity of the source, which is evaluated
accurately, given the distance to {\vw} of 54~pc (\S\ref{section:Intro}).

The other multiple temperature plasma model is {\sc cevmkl}. In {\sc vmcflow}, the temperature
dependence of the differential emission measure ($DEM$) is subject to the temperature dependence of
the volume emissivity of the plasma \citep[so called ``cooling function''; see][for
  example]{1993ApJ...418L..25G}. In {\sc cevmkl}, on the other hand, the $DEM$ is replaced with a
power-law function of the temperature.
\begin{equation}
d (EM)\;\propto \left(\frac{T}{T_{\rm max}}\right)^\alpha\,d(\log T) 
      \;\propto \left(\frac{T}{T_{\rm max}}\right)^{\alpha - 1}\,dT \;\;\;\mbox{for $T < T_{\rm max}$}
\end{equation}

\subsubsection{Reflection model}
\label{sssec:reflection}

Since the boundary layer plasma is formed in contact with the white dwarf surface, a significant
fraction of its X-ray emission is intercepted by the white dwarf and a part of it is reflected off
into space. The height of the boundary layer from the white dwarf surface has been estimated using
the eclipse by the secondary star in the SU~UMa-type DN HT~Cas \citep{1997ApJ...475..812M} to be $< 0.15 R_{\rm WD}$ where $R_{\rm WD}$ is the radius of the white dwarf, and using the intensity of the
iron K$\alpha$ emission line to be $< 0.12 R_{\rm WD}$ in SS~Cyg \citep{2009PASJ...61S..77I}. In the
latter case, a strong fluorescence K$\alpha$ line from neutral iron at 6.40~keV is detected with an
equivalent width of $\simeq$100~eV. Although such a strong line does not seem to be detected from
{\vw}, we have added a plasma emission component that is reflected off the white dwarf to estimate
the intensity of the reflected component. For this, we adopt the {\sc xspec}
model {\sc reflect} \citep{1995MNRAS.273..837M}.

\subsection{Evaluation of the spectra}
\label{ssec:SpecEvaluation}

\begin{figure}
\includegraphics[angle=270,width=1.02\columnwidth]{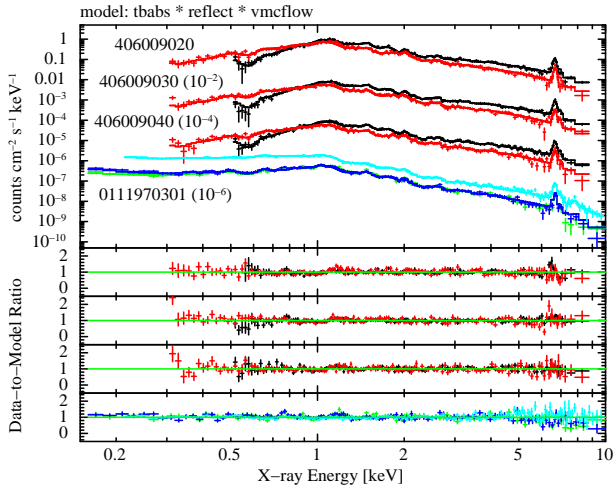}
\caption{Simultaneous fit of the {\sc vmcflow} model to the {\suz} and {\xmm} spectra. Interstellar
  absorption and the reflection are considered with the models {\sc tbabs} (\citep[$N_{\rm H}$ fixed at
  $6\times 10^{17}$~cm$^{-2}$; see ][]{1990ApJ...356..211P} and {\sc reflect}. \label{fig:specfit}}
\end{figure}

We have attempted to fit the {\sc vmcflow} and {\sc cevmkl} models to the {\suz} and {\xmm} spectra
in quiescence. These models include abundances of the major metals as parameters. In order to
make them common among all the spectra, we have tried to fit the four sets of data (three from
{\suz} and one from {\xmm}) simultaneously with the metal abundances being set free to vary but
constrained to have common values. The best-fit continuum parameters with the {\sc vmcflow} and {\sc
  cevmkl} models are summarized in table~\ref{tab:fit_vmcflow} and \ref{tab:fit_cevmkl},
respectively. The resultant elemental abundances with these models are listed in
table~\ref{tab:abundance}.
\begin{table}
\caption{Best-fit continuum parameters with the {\sc vmcflow} model\label{tab:fit_vmcflow}}
\setlength{\tabcolsep}{3pt}
\begin{tabular}{lcccc} \hline
Seq. \# & 406009020 & 406009030 &406009040 & 0111970301 \\ \hline
$N_{\rm H}$ ($10^{17}$ cm$^{-2}$) 
    & \multicolumn{4}{c}{6.0$^\dagger$ (fixed)} \\
$\Omega /2\pi$
    & \multicolumn{4}{c}{1.5$^\ddagger$ (fixed)} \\
$i$ & \multicolumn{4}{c}{60$^\circ$$^\S$ (fixed)} \\
$T_{\rm max}$ (keV)               
    & $8.61^{+0.17}_{-0.26}$ & $8.86^{+0.31}_{-0.22}$ & $8.67^{+0.20}_{-0.22}$ & $6.95^{+0.20}_{-0.19}$ \\
$\dot{M}$ ($10^{-12}M_\odot$yr$^{-1}$) 
    & $2.45^{+0.07}_{-0.05}$ & $1.92^{+0.05}_{-0.06}$ & $2.02^{+0.05}_{-0.05}$ & $1.40^{+0.03}_{-0.03}$ \\ \hline
\multicolumn{5}{l}{Flux$^{\S\S}$ ($10^{-12}$ erg cm$^{-2}$ s$^{-1}$)} \\
~~0.2-12 keV       & 12.63 & 10.18 & 10.49 & 5.81 \\
~~Bolometric$^{||}$ & 14.95 & 12.07 & 12.41 & 6.81 \\ \hline
\multicolumn{5}{l}{Luminosity$^{\S\S}$ ($10^{30}$ erg s$^{-1}$)} \\
~~Bolometric$^{||}$ &  5.21 &  4.21 &  4.33 & 2.37 \\ \hline
$\chi^2$ (d.o.f.)                & \multicolumn{4}{c}{1818.57 (1287)} \\ \hline
\end{tabular}
\begin{minipage}{1.0\columnwidth}
{\sc Note}: All errors are single parameter 90\% confidence level.\\
$\dagger$ Hydrogen column density to {\vw} \citep{1990ApJ...356..211P}.\\
$\ddagger$ Solid angle of cold matter surrounding the boundary layer \citep{2009PASJ...61S..77I}.\\
$\S$ Inclination angle of the reflector (primarily accretion disc), which is regarded as identical
  with the orbital inclination angle \citep{2003A&A...404..301R}.\\
$\S\S$ The reflected component being excluded ($\Omega$ is set equal to zero after the fit converged).\\
$||$ Flux or luminosity in the band 0.04--100~keV.
\end{minipage}
\end{table}
\begin{table}
\caption{Best-fit continuum parameters with the {\sc cevmkl} model\label{tab:fit_cevmkl}}
\setlength{\tabcolsep}{3pt}
\begin{tabular}{lcccc} \hline
Seq. \# & 406009020 & 406009030 &406009040 & 0111970301 \\ \hline
$N_{\rm H}$ ($10^{17}$ cm$^{-2}$)
    & \multicolumn{4}{c}{6.0$^\dagger$ (fixed)} \\
$\Omega /2\pi$
    & \multicolumn{4}{c}{1.5$^\ddagger$ (fixed)} \\
$i$ & \multicolumn{4}{c}{60$^\circ$$^\S$ (fixed)} \\
$T_{\rm max}$ (keV)               
    & $5.06^{+0.24}_{-0.23}$ & $5.88^{+0.39}_{-0.17}$ & $5.77^{+0.16}_{-0.38}$ & $5.32^{+0.23}_{-0.22}$ \\
$\alpha$ 
    & $1.90^{+0.12}_{-0.11}$ & $1.59^{+0.07}_{-0.10}$ & $1.58^{+0.16}_{-0.06}$ & $1.45^{+0.07}_{-0.07}$ \\ \hline
\multicolumn{5}{l}{Flux$^{\S\S}$ ($10^{-12}$ erg cm$^{-2}$ s$^{-1}$)} \\
~~0.2-12 keV       & 12.36 & 10.07 & 10.35 & 5.76 \\
~~Bolometric$^{||}$ & 14.16 & 11.69 & 12.01 & 6.78 \\ \hline
\multicolumn{5}{l}{Luminosity$^{\S\S}$ ($10^{30}$ erg s$^{-1}$)} \\
~~Bolometric$^{||}$ &  4.94 &  4.07 &  4.19 & 2.36 \\ \hline
$\chi^2$ (d.o.f.)                & \multicolumn{4}{c}{1623.98 (1283)} \\ \hline
\end{tabular}
\begin{minipage}{1.0\columnwidth}
{\sc Note}: All errors are single parameter 90\% confidence level.\\
$\dagger$ Hydrogen column density to {\vw} \citep{1990ApJ...356..211P}.\\
$\ddagger$ Solid angle of cold matter surrounding the boundary layer \citep{2009PASJ...61S..77I}.\\
$\S$ Inclination angle of the reflector (primarily accretion disc), which is regarded as identical
  with the orbital inclination angle \citep{2003A&A...404..301R}.\\
$\S\S$ The reflected component being excluded ($\Omega$ is set equal to zero after the fit converged).\\
$||$ Flux or luminosity in the band 0.04--100~keV.
\end{minipage}
\end{table}
\begin{table}
\centerline{
\begin{minipage}{0.7\columnwidth}
\caption{Best-fit elemental abundances with the {\sc vmcflow} and {\sc cevmkl} models \label{tab:abundance}}
\end{minipage}
}
\centerline{
\begin{tabular}{ccc} \hline
   & {\sc vmcflow} & {\sc cevmkl} \\ \hline
O  & $1.02^{+0.10}_{-0.10}$ & $0.93^{+0.10}_{-0.10}$ \\
Mg & $1.56^{+0.18}_{-0.18}$ & $1.02^{+0.19}_{-0.18}$ \\
Si & $1.53^{+0.13}_{-0.13}$ & $1.18^{+0.13}_{-0.13}$ \\
S  & $1.52^{+0.19}_{-0.19}$ & $1.12^{+0.17}_{-0.17}$ \\
Ar & $1.71^{+0.53}_{-0.53}$ & $1.14^{+0.46}_{-0.45}$ \\
Ca & $1.63^{+0.66}_{-0.67}$ & $1.35^{+0.55}_{-0.55}$ \\
Fe & $0.95^{+0.04}_{-0.04}$ & $0.90^{+0.04}_{-0.04}$ \\
Ni & $2.49^{+0.53}_{-0.52}$ & $1.92^{+0.50}_{-0.49}$ \\ \hline
\end{tabular}
}
\centerline{
\begin{minipage}{0.7\columnwidth}
{\sc Note} --- Elemental abundances relative to those of the sun \citep{1989GeCoA..53..197A} are
tabulated.  Abundances of He, C, N, Ne, Na and Al are fixed to unity.
\end{minipage}
}
\end{table}
The result of the fit with the {\sc vmcflow} model is shown in Fig.~\ref{fig:specfit}.

From the $\chi^2$ values, the {\sc cevmkl} model fits better to the observed spectra than the {\sc
  vmcflow} model. The {\sc cevmkl} model has four additional free parameters compared with {\sc
  vmcflow} model. An $F-$test reveals, however, that the addition of the $DEM$ slope parameter ($\alpha$
in eq.(2)) is significant at more than 99\% confidence level. The advantage of the {\sc cevmkl} model
over the {\sc vmcvlow} model is also suggested by \citet{2005ApJ...626..396P} from comprehensive
analysis of DNe data by {\xmm}. 

The solid angle $\Omega/2\pi$ of the reflector subtending over the boundary layer plasma is fixed at
1.5. This is based on the result of the spectral evaluation of SS~Cyg in quiescence
\citep[$\Omega/2\pi = 1.7\pm0.2$; see][]{2009PASJ...61S..77I}. As mentioned in
\S\ref{sssec:reflection}, the spatial extent of the boundary layer is as small as $\sim$0.1 $R_{\rm
  WD}$. If the boundary layer is small enough compared with the white dwarf, the surface of the
white dwarf is viewed from the plasma as an infinite plane covering half of the sky ($\Omega/2\pi =
1$). The accretion disc is further thought to cover half of the remaining sky ($\Omega/2\pi = 0.5$).
In reality, the solid angle should be less than 1.5, because the boundary layer has a finite
size. Nevertheless, the fitting is insensitive to the value of $\Omega/2\pi$, because the reflection
component occupies only a small fraction of the entire spectrum, as will be explained in
\S\ref{ssec:CofReflection} (see Fig.~\ref{fig:reflection}). For the same reason, we have fixed
the inclination of the reflector at the same value as the orbital one 60$^\circ$
\citep{2003A&A...404..301R}.

In the case of using the {\sc cevmkl} model, the maximum temperature of the plasma is nearly
constant, being confined in the narrow range 5--6~keV (table~\ref{tab:fit_cevmkl}). The $DEM$ slope
$\alpha$ of the first {\suz} observation (seq. \#40609020) is, on the other hand, slightly larger
than the other three observations. This means that the spectrum of the first {\suz} observation is
slightly harder than the others. Based on the similarity of $T_{\rm max}$ and $\alpha$, the spectral
shapes of the other three observations resemble one another, although the flux of the {\xmm}
observation (seq. \#0111970301) is roughly half of the second and third {\suz} observations
(seq. \#406009030 and \#406009040). Note that the flux difference between the {\suz} and {\xmm} detectors
is in general less than 10\% \citep{2011A&A...525A..25T,2011PASJ...63S.657I}. Hence, the smaller
flux obtained with {\xmm} by a factor of two is significant. The resultant abundances from the {\sc
  vmcflow} model are in general greater than those from the {\sc cevmkl} model. This is because
$T_{\rm max}$ is larger in the {\sc vmcflow} model; a higher temperature in the keV range reduces
the volume emissivity of the K$\alpha$ emission lines from abundant elements in He-like and
hydrogenic ionization states, and hence requires a larger abundance to explain the observed
intensities of those emission lines. The spectral fits with the {\sc cevmkl} model result in
abundances fairly close to solar values (table~\ref{tab:abundance}), which is consistent with
the results of \citet{2005ApJ...626..396P}.

\subsection{Evaluation of mass accretion rate}
\label{ssec:M-dot}

As explained in \S\ref{ssec:SpecEvaluation}, the model {\sc vmcflow} includes the mass accretion
rate as one of its free parameters. Its best-fit values are listed in table~\ref{tab:M-dot} (the
same as in table~\ref{tab:fit_vmcflow}).
\begin{table*}
\centerline{
\begin{minipage}{0.6\textwidth}
\caption{Mass accretion rates in the unit of $10^{-12}M_\odot$~yr$^{-1}$ obtained from the spectral
  fits. \label{tab:M-dot}}
\end{minipage}
}
\begin{tabular}{llcccc} \hline
Model         & Source    
  & 406009020            & 406009030            & 406009040            & 0111970301 \\ \hline
{\sc vmcvlow} &	model parameter
  & $2.45^{+0.07}_{-0.05}$ & $1.92^{+0.05}_{-0.06}$ & $2.02^{+0.05}_{-0.05}$ & $1.40^{+0.03}_{-0.03}$ \\
              & bolometric flux	
  & $1.67^{+0.03}_{-0.05}$ & $1.35^{+0.04}_{-0.03}$ & $1.39^{+0.03}_{-0.03}$ & $0.86^{+0.03}_{-0.02}$ \\
{\sc cevmkl}  &	bolometric flux
  & $1.58^{+0.03}_{-0.05}$ & $1.30^{+0.04}_{-0.03}$ & $1.34^{+0.03}_{-0.03}$ & $0.76^{+0.03}_{-0.02}$ \\ \hline
\end{tabular}
\end{table*}
In order to confirm these mass accretion rates, as well as to estimate their systematic errors, we
have attempted to calculate them from the observed bolometric fluxes which are listed in
tables~\ref{tab:fit_vmcflow} and \ref{tab:fit_cevmkl} for the {\sc vmcflow} and {\sc cevmkl} models,
respectively. For the conversion to the mass accretion rate, we need the mass and the radius of
the white dwarf. As described in \S1, the mass of the white dwarf has been estimated as $M_{\rm WD} =
0.63\pm0.15 M_\odot$ \citep{1981A&A....97..185S}. With this mass, the radius can be obtained from
the mass-radius relation of \citet{1972ApJ...175..417N} as $R_{\rm WD} =
(8.4^{+1.5}_{-1.2})\times10^8$~cm. As a result, the mass accretion rate can be calculated with the
following equation.
\begin{equation}
\label{eq:M-dot}
  4\pi D^2 f_{\rm bol} \;=\; \frac12\,\frac{GM_{\rm WD}\dot{M}}{R_{\rm WD}},
\end{equation}
where $D = 54$~pc is the distance to the source (\S1), $f_{\rm bol}$ is the bolometric flux listed
in table~\ref{tab:fit_vmcflow} and \ref{tab:fit_cevmkl}. The factor 1/2 on the right-hand side
reflects the fact that half of the accretion energy is lost in the disc before the accreting matter
enters the boundary layer.

The mass accretion rates thus obtained are listed in table~\ref{tab:M-dot}. From this table, they
are found to be smaller than the best-fit values of the {\sc vmcflow} model in general. Some possible
reasons for this will be discussed in \S\ref{ssec:M-dot_Uncertainty}.

\section{Discussion}

\subsection{Continuum model}
\label{ssec:ContModel}
As described in \S\ref{ssec:SpecEvaluation}, the {\sc cevmkl} model provides statistically better
fits to the observed spectra than the {\sc vmcflow} model, even though the former has more free
parameters than the latter. A similar result has been obtained by \citet{2005ApJ...626..396P} from
comprehensive analysis of DNe data by {\xmm}. This may be attributed to the dimensional difference
between galaxy clusters and DNe; the {\sc vmcflow} model presumes spherically symmetric
geometry whereas the accreting gas in DNe is basically in axial symmetry. Since the {\sc
  vmcflow} model was developed for galaxy clusters, the temperature dependence of its $DEM$ can
be different from that of DNe.

\subsection{Contribution of reflection component}
\label{ssec:CofReflection}

\begin{figure}
\centerline{
  \includegraphics[width=0.98\columnwidth]{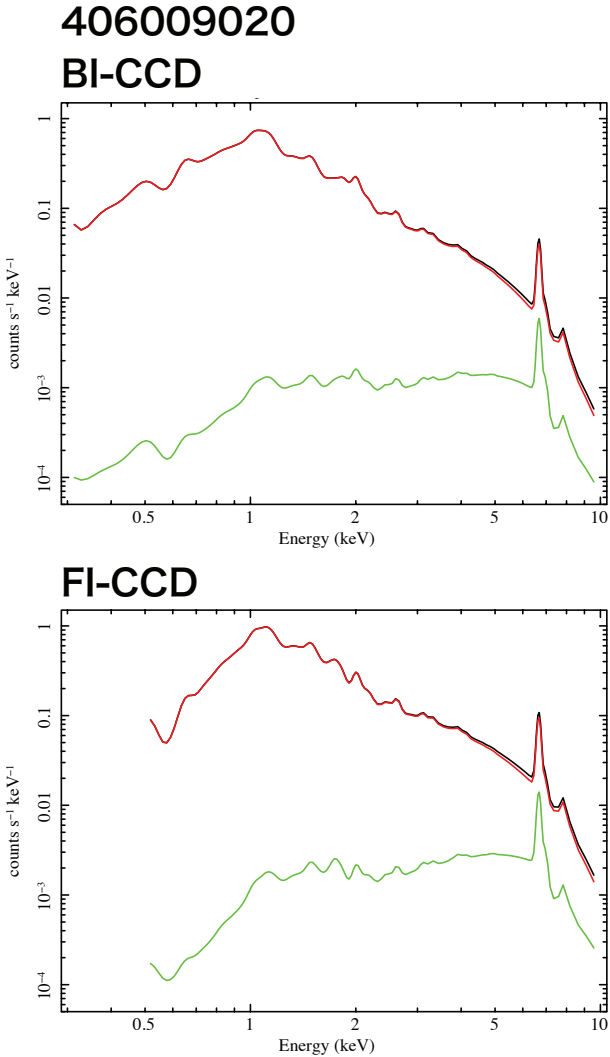}
}
\caption{An example spectral fit to the first {\suz} observation with the {\sc cevmkl} model being
  multiplied with {\sc reflect}, demonstrating contribution of the reflection component with the
  green curve. Black and red curves show models of the total spectrum and the intrinsic plasma
  emission spectrum, respectively. \label{fig:reflection}}
\end{figure}

Fig.~\ref{fig:reflection} shows the best-fit models to the XIS1 (BI-CCD) and XIS0+3 (FI-CCD)
spectra of the first {\suz} observation. 
The intrinsic plasma emission and its reflected component are decomposed and shown in red and
green curves, respectively. In \S\ref{sssec:reflection} we note from geometrical considerations
that there must be a significant contribution to the spectra from the reflected component. As a
matter of fact, a reflected continuum was detected in the {\suz} observation of SS~Cyg in quiescence
\citep{2009PASJ...61S..77I}. In {\vw}, on the other hand, the intensity of the reflected continuum
is very low, approximately $\sim$10\% of the intrinsic plasma emission at 10~keV, and even weaker at lower energies. This difference stems from the fact that the maximum temperature of {\vw} is
much lower than that of SS~Cyg in quiescence ($\simeq$20~keV). In the band below $\sim$10~keV,
photoelectric absorption of the abundant elements dominates the opacity over Thomson
scattering. Hence the reflection component is not conspicuous.

Unlike SS~Cyg, the fluorescent iron emission line at 6.4~keV is almost invisible in {\vw}
(Fig.~\ref{fig:raw_spectra}). This can also be attributed to the lower maximum temperature of the
continuum. The continuum X-ray photons do not have enough energy to ionize K-shell electrons of
iron, whose binding energy is 7.1~keV.

\subsection{Mass accretion rate and its uncertainty}
\label{ssec:M-dot_Uncertainty}

In \S\ref{ssec:M-dot} we have summarized the mass accretion rates during the {\suz} and {\xmm}
observations (table~\ref{tab:M-dot}). The mass accretion rates calculated based on the bolometric
fluxes are in the range 54-70\% of the best-fit $\dot{M}$ of the {\sc vmcflow} model. Possible
reasons for this inconsistency are as follows. First, as described in \S\ref{ssec:ContModel}, the
model parameter $\dot{M}$ in {\sc vmcvlow}, which is based on eq.~(\ref{eq:M-dot}) assumes a 3D
cooling flow geometry whereas the boundary layer is rather close to 2D. This geometrical difference
possibly affects the cooling efficiency of the plasma. As a matter of fact, the {\sc cevmkl} model,
which can adjust the temperature dependence of $DEM$, provides a better fit to the spectra
(\S\ref{ssec:SpecEvaluation}). Second, there remains a significant uncertainty in the white dwarf
mass $M_{\rm WD} = 0.63\pm 0.15M_\odot$ \citep{1981A&A....97..185S}. This can, together with the
associated $R_{\rm WD}$ uncertainty, result in some 50\% uncertainty for $\dot{M}$ calculated from the fluxes. Finally, it is possible that the white dwarf rotates at a
non-negligibly high speed compared to the Keplerian velocity at the white dwarf surface. When the
accreting matter arrives at the boundary layer, it retains kinetic energy of $(R_{\rm WD}\omega_{\rm
  K})^2/2$ per unit mass, where $\omega_{\rm K}$ is the Keplerian angular velocity on the white
dwarf surface in the equatorial plane. Here we neglect the geometrical extent of the boundary layer
for simplicity. If the white dwarf does not rotate at all, all this energy should be released in the
boundary layer for the matter to settle onto the white dwarf. If, on the other hand, the white dwarf
rotates at a finite angular velocity $\omega$, the matter is able to accrete onto the white dwarf by
releasing the energy
\[
\frac12 R_{\rm WD}^2(\omega_{\rm K}^2 - \omega^2) \;=\;
\frac12 \frac{GM_{\rm WD}}{R_{\rm WD}}\left[
   1 - \left( \frac{\omega}{\omega_{\rm K}} \right)^2
   \right]
\]
per unit mass. Consequently, the factor 1/2 in eq.~(\ref{eq:M-dot}) should be replaced by $(1/2) [ 1
  - (\omega/\omega_{\rm K})^2 ]$. The systematic differences apparent in table~\ref{tab:M-dot} can
probably be attributed to these uncertainties and errors.

\subsection{Evolution of the mass accretion rate}

\subsubsection{Evolution of $\dot M$ with supercycle phase}

\begin{figure*}
\centerline{
  \includegraphics[height=0.7\textwidth,angle=270]{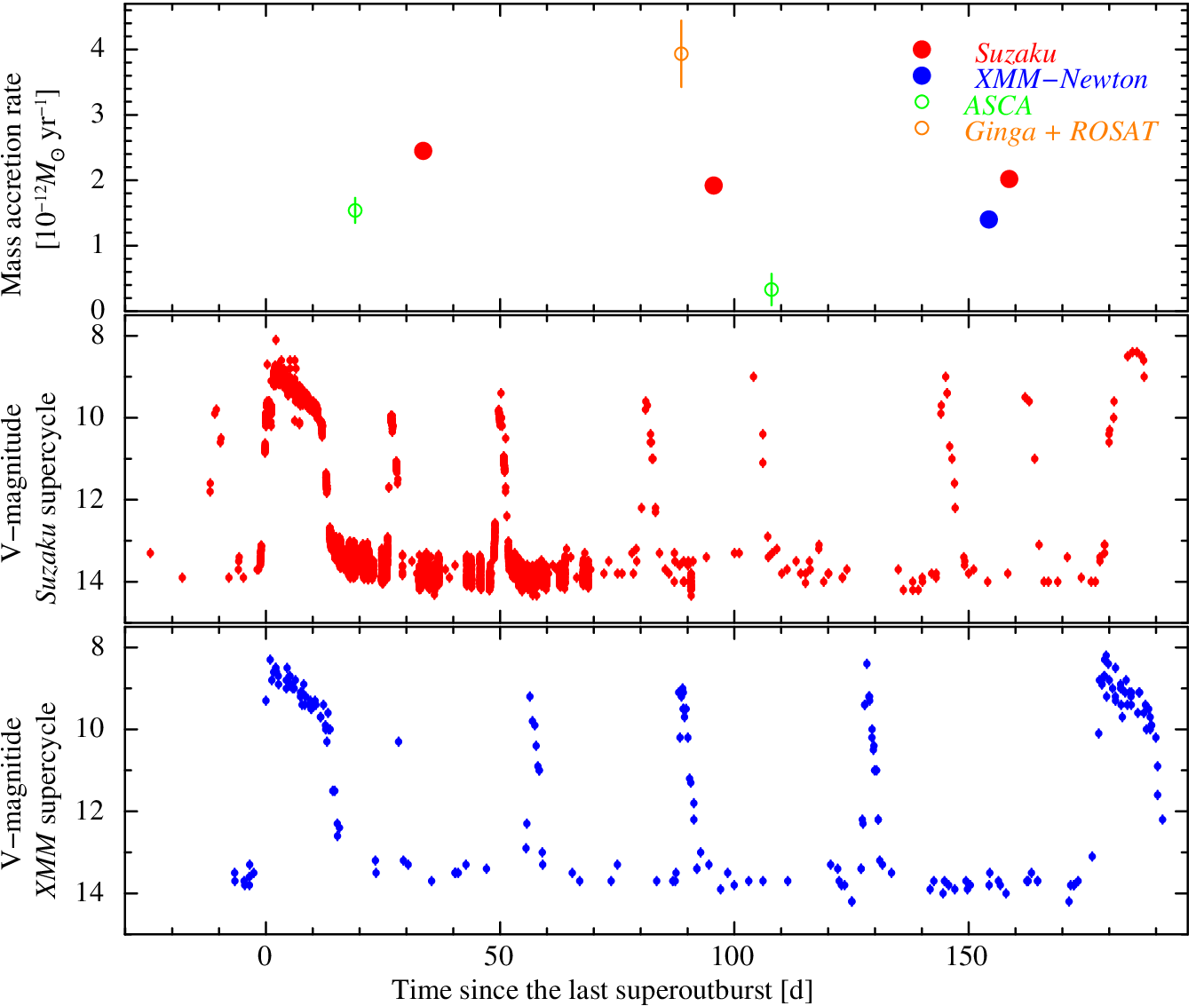}
}
\caption{Time history of the mass accretion rate, the {\sc vmcflow} parameter $\dot M$, since the
  last superoutburst. The filled circles coloured with red and blue in the top panel are from {\suz}
  and {\xmm}, respectively. Open circles coloured with green and orange are from {\asc} observations
  \citep{2005MNRAS.357..626B} and contemporaneous {\gin}+{\ro} observations
  \citep{1996A&A...307..137W}, respectively. The middle and the bottom panels show the V-magnitude
  light curve corresponding to the same supercycles of the {\suz} and {\xmm} observations (the same
  as those in Fig.~\ref{fig:aavso_suzaku} and \ref{fig:aavso_xmm}). \label{fig:Supercycle_M-dot}}
\end{figure*}

In Fig.~\ref{fig:Supercycle_M-dot} we show a time history of the mass accretion rate evaluated in
\S\ref{ssec:SpecEvaluation} and \S\ref{ssec:M-dot_Uncertainty} in comparison to the V-magnitude
light curves drawn since the onset of the last superoutburst before the {\suz} and {\xmm}
observations.
Here we plot the best-fit parameter $\dot M$ of {\sc vmcflow} listed in table~\ref{eq:M-dot}.  Here
we have added a few more data points from past observations. Of the two {\asc} points, the first one
comes from the observation carried out in 1993 November, which is presented in
\citet{2005MNRAS.357..626B}. We have reanalysed the data to find that our bolometric flux coincides
with theirs within 10\%. The other is taken from the observation in 1995 March, which has never been
published. We have processed the data in the same way as the first {\asc} observation, and have
found that {\vw} was extremely faint. We plot the best-fit $\dot M$ of the {\sc vmcflow} model for
both {\asc} datasets. The third data point originates from the {\gin} and {\ro} observation which is
published by \citet{1996A&A...307..137W}. The authors of this paper fitted the contemporaneous
{\gin} and {\ro} spectra with a two component thermal bremsstrahlung model. We have calculated the
bolometric flux from their best-fit model parameters, and derived the mass accretion rate through
eq.~(\ref{eq:M-dot}). In converting this to the {\sc vmcflow} parameter $\dot M$, we refer to the
conversion factor 54\%-70\% from the {\sc vmcflow} parameter $\dot M$ to the flux-based $\dot M$
(\S~\ref{ssec:M-dot_Uncertainty}). We would like to remark, however, that this {\gin}+{\ro}
observation was carried out only a few days after the onset of an outburst, and the hard X-ray
counting rate had been increasing systematically \citep[see Fig. 1 of][]{1996A&A...307..137W}. This
implies that the boundary layer had not reached a steady state.

At first sight of the four data points from the {\suz} and {\xmm} observations, there seems a steady
decline of the mass accretion rate as time passes. Although the first {\suz} observation has the
highest mass accretion rate of the four, the second {\suz} observation gives a similar
mass accretion rate as the third observation, which was carried out some 60 days after
the second. Moreover, the mass accretion rate from the {\xmm} observation, which was performed at
nearly the same epoch as the third {\suz} observation, is smaller than that of the third {\suz} observation by $\sim$30\%. As noted in \S\ref{ssec:SpecEvaluation}, disagreement of fluxes measured
with {\suz} and {\xmm} detectors is less than 10\%
\citep{2011A&A...525A..25T,2011PASJ...63S.657I}. Hence this flux disagreement is real. Although the
three {\suz} observations and the {\xmm} observation belong to different supercycles, we
conclude that there is no clear trend in the mass accretion rate as a function of the time since the
eruption of a superoutburst. This conclusion holds, or is even reinforced, with the {\asc} and
{\gin} + {\ro} data points being included.

We remark that, in SU~UMa type DNe, the total mass of the accretion disc increases between two
superoutbursts \citep{1996PASP..108...39O}. As the disc accumulates mass, its surface density is
expected to increase as time passes, and hence the mass accretion rate also increases. Such a 
systematic increase of the mass accretion rate is not evident from the current {\suz} and {\xmm}
observations.

\subsubsection{Evolution of $\dot M$ in the outburst cycle}

\begin{figure}
\centerline{
  \includegraphics[height=\columnwidth,angle=270]{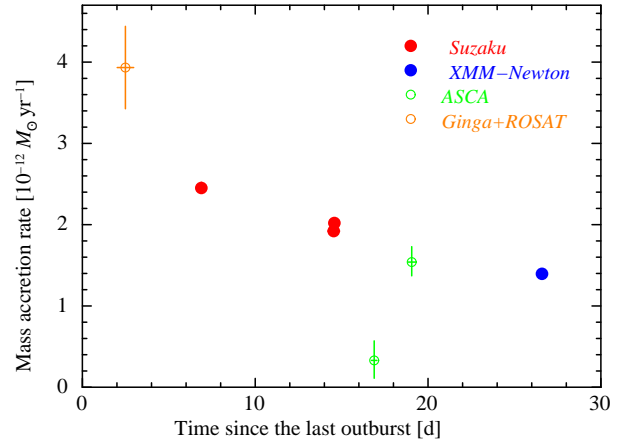}
}
\caption{Evolution of the mass accretion rate as a function of the time since the last outburst. The
  symbols of the data points are the same as those in
  Fig.~\ref{fig:Supercycle_M-dot}. \label{fig:Burstcycle_M-dot}}
\end{figure}

In Fig.~\ref{fig:Burstcycle_M-dot}, we have plotted the mass accretion rate as a function of the
time since the last outburst (including the superoutburst).
In contrast to Fig.~\ref{fig:Supercycle_M-dot}, this shows a clear decreasing trend of the mass
accretion rate, except for the extremely low mass accretion rate in the second {\asc} observation.
There are at least a couple of previous works that suggest decline of the hard X-ray flux during
optical quiescence. \citet{2004ApJ...601.1100M} analysed four sets of {\it RXTE} PCA data of
SS~Cyg, each of which encompass $\sim$2 weeks to a few months. They found that the X-ray counting
rate in the band 1.3--12.2~keV declines at a rate of 1.3\% d$^{-1}$, accumulating 40\% decrease in
31 days since the instance when SS Cyg gets fainter than $m_{\rm V} =
11.7$~mag. \citet{2010MNRAS.402.1816C} found that the X-ray counting rate of SU~UMa in quiescence in
the band 2.0-18.5~keV observed with the {\it RXTE} PCA decreases at a rate of 0.09
counts~s$^{-1}$~d$^{-1}$ from 5.8 counts s$^{-1}$, which implies $-1.6$\%~d$^{-1}$, similar to
SS~Cyg. The decline of the mass accretion rate in the current dataset of {\vw} is 43\% in
$\sim$20~d, implying the rate of $-2.2$\%~d$^{-1}$. This is of the same order as the previous two
works.

The authors of these two previous works mentioned that the global decline of the mass accretion rate
within the boundary layer during the quiescent phase (inter-outburst period) conflicts with the
expectation of the disc instability model, which predicts increasing accretion rate during
quiescence \citep[e.g.][]{2001NewAR..45..449L}. One may invoke retardation of mass transfer through
the disc. \citet{1985PASJ...37....1M} showed in their disk instability simulation that the outburst
is triggered by a heating wave that is initiated in the outer part of the disc and travels inward. The
expected delay of an outburst in the EUV band is really detected in SS~Cyg;
\citet{2003MNRAS.345...49W} found that the rise of the EUV flux is retarded from the onset of the
optical outburst by 1.5-2~d. \citet{2010MNRAS.402.1816C} found that in SU~UMa the hard X-ray
suppression, which is expected to occur simultaneously with the EUV outburst, is retarded by 0.57~d
on the average from the onset of the optical outburst. These results indicated that it takes $\sim$1
day for the heating wave to reach the boundary layer from the optically bright part of the
disc. Since the quiescent disc is expected to have much lower density, the time necessary for physical
quantities to propagate through the disc should be much longer. 

According to Fig.~1 of \citet{1996A&A...307..137W} (and also Fig.~\ref{fig:aavso_suzaku} and
\ref{fig:aavso_xmm}), however, a normal optical outburst continues only $\sim$3 days. The end of the
outburst corresponds to the epoch when the cooling wave passes the optical disc and arrives at the
entrance of the boundary layer, after which the mass accretion rate should resume as per the
standard disc instability model. The increase of the hard X-ray flux \citep[again Fig.~1
  of][]{1996A&A...307..137W}, which is direct evidence of a decline of the mass accretion rate in the
boundary layer, continues at least until $\sim$6 days after the onset of the optical
outburst. This seems to conflict with the disc instability model. One may consider
  that Fig.~1 of \citet{1996A&A...307..137W} did not detect the end of the outburst due to the optical
  flux limit, yet it is hard to believe that it takes $\sim$30~d (Fig.~\ref{fig:Burstcycle_M-dot})
  for the cooling wave to arrive at the boundary layer, since 30~d is comparable to the average
  time of burst repetition (Fig.~\ref{fig:Supercycle_M-dot}).

To reconcile the mass accretion rate decline in quiescence with the standard disc instability
model, one may presume that the optically thick disc survives close to the orbital plane even in
the optically-quiescent state, being covered with an optically thin boundary layer. Since there must be
some density gradient in the disc in the direction perpendicular to the orbital plane, such a
geometry may be possible. As the mass accretion rate increases, this optically thick boundary layer
grows and consumes a larger fraction of the accretion energy, resulting in the apparent decline of
$\dot M$ in the hard X-ray band. The emission from such a disc is expected to
appear in the EUV band, being somewhat Comptonized by the overlying optically thin hot plasma. Such a
spectral component, however, has not been detected so far; we should confess that such an optically thick
component of the boundary layer in quiescence is just speculation at this moment.

On the other hand, the standard disk instability model \citep[][for example]{2001NewAR..45..449L}
may need to be modified. Although the model predicts an increase of the mass accretion rate through the
disc in quiescence, the optical magnitude seems fairly constant throughout the quiescence phase (see
Fig.~\ref{fig:aavso_suzaku} and \ref{fig:aavso_xmm}). The optical flux originates not only from the
accretion disc but also from the white dwarf and the secondary star. Of them, the emission from the
white dwarf may decline during quiescence due to cooling after the outburst. It is, however,
difficult to imagine that the increase of the disc emission and the decrease of the white dwarf
emission just compensate each other. We also believe there is no DN that shows a steady decline of the
V-magnitude (flux increase) during optical quiescence.

Finally, we briefly mention the extremely faint state of {\vw} observed with {\asc} in March
1995. {\vw} was so faint and the mass accretion rate at that time was much lower than the declining
trend shown in Fig.~\ref{fig:Burstcycle_M-dot}. Although the spectral parameters are not constrained
very well, the {\asc} spectrum is fitted with the {\sc vmcflow} model with $T_{\rm max} =
11.5^{+3.5}_{-3.1}$~keV. The spectrum seems slightly harder than the other quiescence spectra
presented in this paper. This can be attributed to a lower mass accretion rate, resulting in lower
cooling efficiency of the plasma. The reason for {\vw} to fall into such a low state should be pursued
in the future.

\section{Conclusion}

In order to elucidate time variation of the mass accretion rate of DNe in the optically-quiescent
phase, we have analysed a series of {\vw} data taken with {\suz}. As complement, we have added one
dataset from {\xmm}. The observed spectra in the band 0.2-10~keV are moderately well represented by
multiple temperature thermal plasma emission models ({\sc vmcflow} and {\sc cevmkl} models in {\sc
  xspec}) with a maximum temperature of 5-9~keV. The bolometric luminosity is $(2.4-5.2)\times
10^{30}$ erg~s$^{-1}$. From the evaluation of the spectra, we have derived the mass accretion
rates. According to the theoretical prediction of the disc behaviour during a supercycle of
SU~UMa type DNe \citep{1996PASP..108...39O}, the disc accumulates mass throughout the entire
supercycle, and the mass accretion rate through the disc is expected to increase. The measured mass
accretion rate, however, does not show any clear trend with the supercycle phase. If we plot the
mass accretion rate, on the other hand, against the elapsed time since the last outburst (including
the superoutburst), we have found a systematic declining trend with a rate of
$-2.2$\%~d$^{-1}$. This rate of decline is comparable to the other two DNe: SS~Cyg
\citep[$-1.3$\%~d$^{-1}$,][]{2004ApJ...601.1100M} and SU~UMa
\citep[$-1.6$\%~d$^{-1}$,][]{2010MNRAS.402.1816C}. This trend holds, or is even reinforced by adding
a few datasets from {\gin}, {\ro} and {\asc}. The decrease of the mass accretion rate through the
boundary layer during the optically-quiescent state is, however, hard to reconcile with the
standard disc instability model \citep[e.g.][]{2001NewAR..45..449L} in which the mass accretion rate
is predicted to increase throughout the quiescence phase. This discrepancy may be resolved if one
may speculate that an optically thick disc survives in the orbital plane of the boundary layer even
during the optically-quiescent state, consuming a larger fraction of the accretion energy than the
optically thin hot plasma as the mass accretion rate increases. Alternatively, the standard disc
instability model may need to be modified. We obviously need further observation and theoretical
consideration to finally resolve this discrepancy.

\section*{Acknowledgements}

The authors are grateful to the anonymous referee for useful suggestions and comments. They would
like to express special gratitude to Dr. Koji Mukai for his useful information and constructive
discussion. They are also grateful to Prof. Takaya Ohashi and Dr. Magnus Axelsson for his invaluable comments to the initial
draft of this paper.

This work has made use of data from the European Space Agency (ESA) mission {\it
    Gaia}\footnote{\url{https://www.cosmos.esa.int/gaia}}, processed by the {\it Gaia} Data
  Processing and Analysis Consortium (DPAC)\footnote{
  \url{https://www.cosmos.esa.int/web/gaia/dpac/consortium}}. Funding for the DPAC has been provided
  by national institutions, in particular the institutions participating in the {\it Gaia}
  Multilateral Agreement.



\bibliographystyle{mnras}
\bibliography{vwhyi} 

%
%
%
%

\bsp	
\label{lastpage}
\end{document}